\documentstyle[astrobib,psfig,url]{mn-ab}
\def\apj{ApJ}
\def\apjl{ApJL}
\def\apjs{ApJS}
\def\mnras{MNRAS}
\def\aap{A\&A}
\def\araa{ARA\&A}
\def\nat{Nat}

\def\aj{AJ}
\def\pasj{PASJ}

\def\pasp{PASP}
\def\aaps{Acta Astrophys. Sin.}

\def\ref{\par\noindent\hang}
\def\spose#1{\hbox to 0pt{#1\hss}}
\def\approxlt{\mathrel{\spose{\lower 3pt\hbox{$\sim$}}
        \raise 2.0pt\hbox{$$<$$}}}
\def\approxgt{\mathrel{\spose{\lower 3pt\hbox{$\sim$}}
        \raise 2.0pt\hbox{$>$}}}

\def\multleft#1{\hbox to size{\vbox {\halign {\lft{##}\cr #1}}\hfill}\par}
\def\multright#1{\hbox to size{\vbox {\halign {\rt{##}\cr #1}}\hfill}\par}

\def\today{\ifcase\month\or January\or February\or March\or April\or May\or
      June\or July\or August\or September\or October\or November\or December\fi
      \space\number\day, \number\year}
\def\s{\hbox{\phantom{5}}}      

\def\boxit#1{\vbox{\hrule\hbox{\vrule\kern3pt\vbox{\kern3pt
          #1 \kern3pt}\kern3pt\vrule}\hrule}}

\def\cm{{\rm\thinspace cm\thinspace}}

\def\erg{{\rm\thinspace erg\thinspace}}
\def\eV{{\rm\thinspace eV\thinspace}}

\def\K{{\rm\thinspace K\thinspace}}
\def\keV{{\rm\thinspace keV\thinspace}}
\def\km{{\rm\thinspace km\thinspace}}

\def\s{{\rm\thinspace s\thinspace}}

\def\ergcmps{\hbox{$\erg\cm\ps\,$}}

\def\ergpcmsqps{\hbox{$\erg\cm^{-2}\s^{-1}\,$}}

\def\ergps{\hbox{$\erg\s^{-1}\,$}}

\def\kmps{\hbox{$\km\s^{-1}\,$}}

\def\pcm{\hbox{$\cm^{-3}\,$}}
\def\pcmsq{\hbox{$\cm^{-2}\,$}}

\def\ps{\hbox{$\s^{-1}\,$}}

\def\asca{\hbox{\em ASCA\thinspace}}
\def\chandra{\hbox{\em Chandra\thinspace}}

\input{psfig.sty}

\begin{document}
\hsize=6truein

\title{Weighing black holes with warm absorbers}
\author[R. Morales and A.C. Fabian]{
  R. Morales\thanks{E-mail: rm@ast.cam.ac.uk} and A. C.
Fabian\\
  Institute of Astronomy, Madingley Road, Cambridge CB3 0HA}
\maketitle

\begin{abstract} 
  We present a new technique for determining an upper limit for the
  mass of the black hole in active galactic nuclei showing warm
  absorption features. The method relies on the balance of radiative
  and gravitational forces acting on outflowing warm absorber clouds.
  It has been applied to 6 objects: five Seyfert 1 galaxies: IC
  4329a, MCG$-$6-30-15, NGC 3516, NGC 4051 and NGC 5548; and one
  radio-quiet quasar: MR 2251-178. We discuss our result in
  comparison with other methods. The procedure could also be 
  applied to any other radiatively driven optically thin outflow in
  which the spectral band covering the major absorption is directly
  observed. 
\end{abstract}

\begin{keywords} galaxies: active  $-$ galaxies: Seyfert $-$ X-rays: galaxies.
\end{keywords}

\section{INTRODUCTION}
It is the current paradigm that the fundamental power source of all
Active Galactic Nuclei (AGN) is accretion onto a supermassive black
hole (SMBH) (see \citeNP{Lynd69} and \citeNP{Rees84}). For many years,
theorists and observers have attempted to prove the existence of
SMBHs, inferring their presence by studying how they affect the
surrounding environment (\citeNP{Korm95} and \citeNP{Rich98} review
the SMBH search).  A variety of techniques and arguments has been
developed to estimate the mass of the central engine in AGN
\cite{Ho00}. This has included both theoretical and observational
methods such as: fitting the spectra of accretion discs; high
efficiency and Eddington limit arguments; stellar dynamical searches;
kinematics of radio masers; X-ray variability and virial masses from
optical variability, which includes reverberation mapping and
photoionization model techniques.

In this paper we propose the use of warm absorbers (WAs) to determine
an upper limit to the BH mass. WAs are significant quantities of
partially-ionized, optically-thin material found mainly in the
vicinity of Type 1 Seyfert galaxies. This gas is most readily
observable when it lies along line-of-sight to the central X-ray
source: the observed X-ray spectrum then displays well-defined K-shell
absorption edges. \citeN{hal84} was the first to postulate the
presence of such material based on the unusual X-ray spectrum of the
QSO MR 2251-178 from the \textit{Einstein} observatory. The clearest
demonstration of the presence of warm absorbing material comes from
spectroscopic observations of the OVII and OVIII K-shell absorption
edges at 0.74 and 0.87 \keV, respectively. These edges were first
found by the \textit{ROSAT} PSPC \cite{Nand92}. The spectral
capabilities of \textit{ASCA} allowed for detailed studies of such
edges (e.g. \shortciteNP{fab94}, \shortciteNP{ota96},
\shortciteNP{Brand96}, \shortciteNP{Mat97}, \shortciteNP{reyno97},
\shortciteNP{Leig97} and \shortciteNP{Geor98}). WAs are common, being
detected in half of the sources (\citeNP{reyno97},
\shortciteNP{Geor98}). Several different models have been suggested
concerning the nature and location of the WA.  Among others, these
include the very detailed studies by \shortciteN{Krol95} and
\shortciteN{Netz96}. The reason for the large variety of models is
that not all physical properties of the WA can be directly determined
from X-ray spectral fits, but only certain combinations of them.

In addition to the WAs, associated UV and optical absorption features
are also typically found in AGN spectra (see \shortciteNP{Cren99} and
references therein for an overview). A common property shared by
almost all associated absorption features is that any velocity shift
is always to the \textit{blue} of the systemic velocity, implying that
the absorbing material is outflowing.  \shortciteN{Cren99} found a
one-to-one correspondence between those objects showing UV absorption
and those presenting X-ray WAs. This implies the two phenomena are
likely to be physically related.  However, since WAs generally reveal
higher ionization states, and higher column densities, they can not be
the same in general. A very plausible connection is a dynamical one in
which both UV absorbers and X-ray WAs are outflowing. The blue-shifts
in X-ray absorption lines have recently been revealed by
\textit{Chandra} observations (e.g.  \citeNP{Kaas00},
\citeNP{Kasp00a}, \citeNP{Coll01}), presenting typical velocities of
several hundred to several thousand \kmps. One unresolved issue is
whether the outflowing gas is smoothly distributed (i.e. winds) or
resides in discrete clouds.  \shortciteN{Krol95} show that the warm
absorption seen in many X-ray spectra of AGN may be produced by X-ray
heated winds.  \shortciteN{Reyn95} also sketch some ideas relating the
WA to a radiatively driven outflow. In contrast to the wind model,
\shortciteN{Chel01} show that accelerations to velocities of a few
hundred to a few thousand \kmps is a natural consequence of the AGN
environment when radiation pressure acts on external pressure-confined
highly ionized clouds (responsible for the WA features). For the
present study, we will adopt the latter model in which WA clouds are
radiatively accelerated by the central ionizing continuum.

This work describes a new method for obtaining an upper limit on an
AGN BH mass, by means of a comparison between the radiative and
gravitational forces acting on its WA clouds. Section 2 presents the
30 observations of the 6 AGNs sample to which our technique has been
applied. The description of our method, a spectral analysis of the
\textit{ASCA} data and the results and comparison with other BH mass
estimates by different techniques are addressed in Section 3.  Section
4 and 5 analyse the influence of the underlying continuum on our
estimate and further terms in the equation of motion, respectively.
Finally, in Section 6 our conclusions are summarized. Throughout this
paper, unless otherwise stated, errors on physical quantities are
quoted at the 1$\sigma$ level for one interesting parameter.
\begin{table}
\medskip
\begin{center}
\begin{tabular}{lcccc} \hline\hline
Source & Type & Redshift & Galactic $N_H$ & Date of\\
   &  & z & ($10^{20}$ $cm^{-2}$) & Observation\\
\hline
IC 4329a & Sy1 & 0.016 & $4.55^a$ & 93/08/15\\
 & & & & 97/08/07\\
 & & & & 97/08/10\\
 & & & & 97/08/12\\
 & & & & 97/08/15\\
\hline
MCG$-$6-30-15 & Sy1 & 0.008 & $4.06^a$ & 93/07/09\\
 & & & & 93/07/31\\
 & & & & 94/07/23\\
 & & & & 97/08/03\\
 & & & & 97/08/07\\
\hline
MR 2251-178 & RQQ & 0.068 & 2.8 & 93/11/16\\
 & & & & 93/12/07\\
 & & & & 93/12/14\\
 & & & & 93/12/19\\
 & & & & 93/12/24\\
\hline
NGC3516 & Sy1 & 0.009 & 3.4 & 94/04/02\\
 & & & & 95/03/11\\
 & & & & 95/03/12\\
\hline
NGC4051 & Sy1 & 0.002 & $1.31^a$ & 93/04/25\\
 & & & & 94/06/07\\
\hline
NGC5548 & Sy1 & 0.017 & 1.7 & 93/07/27\\
 & & & & 96/06/27\\
 & & & & 96/06/29\\
 & & & & 96/07/01\\
 & & & & 96/07/03\\
 & & & & 96/07/04\\
 & & & & 98/06/15\\
 & & & & 98/06/20\\
 & & & & 98/07/07\\
 & & & & 99/01/19\\
\hline
\medskip
\end{tabular}
\end{center}
\caption{The AGN sample. Column 1 gives the common source name. Column
   2 indicates the type of nuclear activity (Sy1 $\equiv$ Seyfert 1
   galaxy; RQQ $\equiv$ radio-quiet quasar). Column 3 give the
   redshift of the sources (Veron-Cetty $\&$ Veron 1993). Column 4
   gives the Galactic HI column density towards the source as
   determined by 21-cm measurements [``a'' indicates an accurate value
   obtained from Elvis, Wilkes $\&$ Lockman (1989); otherwise quoted
   values are interpolations from the measurements of Stark et
   al. (1992)]. Column 5 gives the year/month/day of the \asca observation used
   (see the Tartarus Database for a detailed description of each observation).}
\end{table}

\section{THE DATA}

We have studied a sample of six AGNs consisting of five Seyfert 1
galaxies (IC 4329a, MCG$-$6-30-15, NGC 3516, NGC 4051 and NGC 5548)
and one radio-quiet quasar (MR 2251-178), which cover a range of
almost 4 orders of magnitude in X-ray luminosity. These objects are a
subsample of those presenting warm absorption features in the 24 Type
1 AGN sample of Reynolds (1997). With the exception of MR 2251-178, the
other five Seyfert galaxies are also present in the
\shortciteN{Geor98} survey. The WAs of all these sources have a very
rich observational history \cite{Komo99} and their host AGNs are
amongst the brightest ones in the nearby Universe ($z\geq 0.07$). The
spectral files have been downloaded from the Tartarus
Database (\url{http://tartarus.gsfc.nasa.gov/}) and the 30
observations analysed are listed in Table
1\nocite{Vero93}\nocite{Elvi89}\nocite{Star92}.

\section{UPPER BLACK HOLE MASS ESTIMATE}

\subsection{Description of the method}

The following assumptions have been made:
\begin{itemize}
 \item the WA, responsible for the absorption features around 1 \keV, is
   present in the form of clouds.
 \item these clouds are pushed radiatively outwards.
 \item the clouds motion is entirely radial.  
\end{itemize}

The radial equation of motion for these WA clouds is
\begin{equation}
\frac{AL_{\rm abs}}{4\pi r^2c}-\frac{GM_{\rm BH}m_{\rm wa}}{r^2}-\rho _{\rm o}v^2A=m_{\rm wa}a
\end{equation}
where: $A$ is the cross-section presented by the WA clouds; $L_{\rm
  abs}$ is the luminosity absorbed by the WA cloud; $r$ is the
distance from the ionizing source to the WA cloud; $c$ is the velocity
of light; $G$ is the gravitational constant; $M_{\rm BH}$ is the black
hole mass; $m_{\rm wa}$ is the mass of a WA cloud; $\rho_{\rm o}$ is
the density of the medium through which the WA is moving; $v$ is the
velocity of the WA clouds with respect to that medium and $a$ is the
acceleration experienced by the WA cloud.

The first term on the left hand side is the outward force due to
radiation pressure, the second term is the restoring force due to
gravity, and the third term describes any drag force (i.e. ram
pressure) experienced by the WA when moving through an intercloud
medium. If the drag force and the acceleration term are negligible, a
comparison between the radiation and gravitation terms leads to an
expression for the BH mass:
\begin{equation}
M_{\rm BH}=\frac{L_{\rm abs}}{4\pi cGm_{\rm p}N_{\rm wa}}
\end{equation}
where $m_{\rm p}$ is the proton rest-mass and $N_{\rm wa}$ is the
column density of the WA clouds. If these terms are non negligible and
positive, then the above expression becomes an upper limit for $M_{\rm
  BH}$. We shall treat it as such in the rest of the paper and discuss
further the drag and acceleration terms in Section 5.

The only two unknowns in Eqn. 2 are $L_{\rm abs}$ and $N_{\rm wa}$.
$N_{\rm wa}$ is calculated as a parameter in our spectral fits of a
single ionization parameter model presented in the following
subsection. As already seen in \citeN{reyno97} and
\shortciteN{Geor98}, the data can often be well fit by such a model
which allows for accurate estimates of the ionization parameter and
column densities. The calculation of $L_{abs}$ has been performed
using the fact that the photoionized plasma causing the edges is
optically thin along the line of sight.  Therefore, the expression for
the radiation pressure, $P_{\rm rad}$, can be written as
\begin{equation}
P_{\rm rad}=\frac{L_{\rm abs}}{4\pi r^2c}=\frac{1}{4\pi cr^2} \int^{\nu _{\rm max}}_{\nu _{\rm min}}(L(\nu)-L_{\rm obs}(\nu ))d\nu
\end{equation}
where: $L(\nu)$ is the luminosity per unit frequency from the ionizing
source (which we assume to be a power law of photon index $\Gamma$)
and $L_{\rm obs}(\nu )$ is the luminosity per unit frequency observed
after $L(\nu)$ has passed through the WA cloud. The photoionization
models can predict both the intensity of the incident and transmitted
continuum allowing the calculation of the above integral (see
Section 3.2.3). We have chosen the limits of our integration to
be: $\nu _{\rm min}=1$ \eV and $\nu _{\rm max}=20$ \keV.

\subsection{Spectral analysis}
The data have been fitted using two different models. The aim for this
multiple analysis has been firstly, to determine phenomenologically
the underlying continuum by mimicking the WA features by the two
photoionization absorption edges at the OVII and OVIII rest-frame
energies (Sections 3.2.1 and 3.2.2); and then to calculate the
radiation pressure acting on the WA (Section 3.2.3).  Data from all
four \asca instruments were fitted simultaneously. SIS data between
0.5--4.0 \keV and GIS data between 0.8--4.0 \keV were used.  The
spectra were then fitted using the standard $\chi^2$ minimization
technique implemented in version 11.0.1 of the \textsc{xspec} spectral
fitting package \cite{arna96}.

\subsubsection{Phenomenological model}

In order to determine the underlying continuum hidden by various
spectral features displayed by the data, the spectrum of each object
was fitted by a spectral model consisting of the following components:
\begin{itemize}
 \item a power-law representing the primary continuum (photon index
   $\Gamma$)
 \item two absorption edges with rest-frame threshold energies of 0.74
   \keV and 0.87 \keV representing OVII and OVIII K-shell absorption
   (with maximum optical depths $\tau_{\rm O7}$ and $\tau_{\rm O8}$)
   respectively. These two edges provide, to a good approximation, a
   description of the effects of the warm absorber over the ASCA band.
 \item intrinsic absorption by a column density $N_{\rm H}$ of neutral
   matter in the rest frame of the source. This absorbing material is
   assumed to have cosmic abundances. 
 \item Galactic absorption (at $z=0$) fixed at the level determined by
   the HI 21-cm measurements reported in Table 1, column 4.
\end{itemize}
Free parameters in the fit were $\Gamma$, $\tau_{\rm O7}$, $\tau_{\rm O8}$ and
$N_{\rm H}$. The normalizations of the model for each of the four
instruments were also left as free parameters. This allows for the
known $\sim 10$--$20$ per cent discrepancies between the normalizations
of different instruments (see
\url{http://heasarc.gsfc.nasa.gov/docs/asca/watchout.html} for a discussion
of the decrease in efficiency of SIS0 and SIS1 below 1 \keV since
approximately late 1994). Table 2 presents a summary of the $\chi^2$
fitting.

As noted in Reynolds (1997)\nocite{reyno97}, fixing the energies of the
two absorption edges at the physical rest-frame energies of the
K-shell absorption edges of OVII and OVIII, is the most robust to the
various unmodeled spectral complexities such as recombination
line/continuum emission, resonance absorption lines and other
absorption edges. These complexities would require many additional
degrees of freedom to model (such as the covering fraction of the warm
material, its density, the chemical abundances and the velocity
structure) and would lead to an over modeling of the data for all but
the very highest quality \asca data. In fitting a simple two-edge
model, these complexities could lead to false shifts in the threshold
edge of the edges (e.g. the existence of the Ne IX edge may falsely
imply a small blueshift of the OVIII edge when a two-edge model is
fitted with the edge threshold energies as free parameters).
\begin{table*}
\begin{minipage}{170mm}
\medskip
\begin{center}
\begin{tabular}{lccccccc} \hline\hline
Source & Date of & Photon index & & & Intrinsic $N_{\rm H}$ &
 $\chi^2$/dof & Red. $\chi^2$\\
 & Observation & $\Gamma$ & $\tau_{\rm O7}$ & $\tau_{\rm O8}$ & ($10^{22}$ \pcmsq) & &\\
\hline
IC4329a & 93/08/15 & $1.85^{ +0.01}_{ -0.01}$ & $0.67^{ +0.06}_{
 -0.06}$ & $0.06^{ +0.04}_{ -0.04}$ & $0.30^{ +0.01}_{
 -0.01}$ & 1036.6/776 & 1.34\\
 & 97/08/07 & $1.90^{ +0.01}_{ -0.01}$ & $0.11^{ +0.07}_{
 -0.07}$ & $0.00^{+0.01}_{ -0.00}$ & $0.44^{+0.01}_{
 -0.01}$ & 1328.5/1007 &1.32\\
 & 97/08/10 & $1.79^{ +0.01}_{-0.02}$ & $0.45^{+0.06}_{
 -0.08}$ & $0.00^{+0.02}_{-0.00}$ & $0.34^{+0.01}_{
 -0.01}$ & 825.5/776  & 1.06\\
 & 97/08/12 & $1.91^{+0.02}_{-0.02}$ & $0.28^{+0.08}_{
 -0.08}$ & $0.00^{+0.02}_{-0.00}$ & $0.42^{+0.01}_{
 -0.01}$ & 896.2/776 & 1.15\\
 & 97/08/15 & $1.91^{+0.02}_{-0.02}$ & $0.09^{ +0.10}_{
 -0.10}$ & $0.04^{+0.04}_{-0.04}$ & $0.47^{ +0.02}_{
 -0.02}$ & 1308.2/1009 & 1.30\\
\hline
MCG$-$6-30-15 & 93/07/09 & $2.01^{ +0.01}_{ -0.01}$ & $0.65^{
 +0.03}_{-0.03}$ & $0.23^{ +0.02}_{-0.02}$ & $0.00^{
 +0.00}_{-0.00}$ & 1203.7/775 & 1.55\\
 & 93/07/31 & $1.88^{+0.01}_{-0.01}$ & $0.70^{+0.04}_{
 -0.04}$ & $0.49^{ +0.03}_{ -0.03}$ & $0.00^{+0.00}_{
 -0.00}$ & 1110.9/772 & 1.44\\
 & 94/07/23 & $1.88^{ +0.01}_{ -0.01}$ & $0.67^{ +0.01}_{
 -0.01}$ & $0.12^{ +0.01}_{ -0.01}$ & $0.011^{ +0.002}_{
 -0.002}$ & 2861.5/1013 & 2.82\\
 & 97/08/03 & $1.80^{ +0.01}_{ -0.01}$ & $0.91^{ +0.03}_{
 -0.03}$ & $0.07^{ +0.03}_{ -0.03}$ & $0.032^{ +0.004}_{
 -0.004}$ & 1972.6/1013 & 1.95\\
 & 97/08/07 & $1.77^{ +0.01}_{ -0.01}$ & $0.73^{ +0.03}_{
 -0.03}$ & $0.15^{ +0.03}_{ -0.03}$ & $0.036^{ +0.003}_{
 -0.004}$ & 1649.4/1015 & 1.63\\
\hline
MR 2251-178 & 94/04/02 & $1.81^{ +0.02}_{ -0.02}$ & $0.40^{
 +0.07}_{ -0.07}$ & $0.17^{ +0.05}_{ -0.06}$ & $0.000^{
 +0.003}_{ -0.000}$ & 829.3/823 & 1.01\\
 & 93/12/07 & $1.75^{ +0.02}_{ -0.02}$ & $0.41^{ +0.06}_{
 -0.05}$ & $0.10^{ +0.04}_{ -0.04}$ & $0.000^{  +0.004}_{
 -0.000}$ & 821.7/867 & 0.95\\
 & 93/12/14 & $1.67^{ +0.03}_{ -0.02}$ & $0.40^{ +0.08}_{
 -0.09}$ & $0.09^{ +0.07}_{ -0.05}$ & $0.000^{ +0.007
 }_{-0.000}$ & 768.0/750 & 1.02\\
 & 93/12/19 & $1.63^{ +0.04}_{ -0.04}$ & $0.5^{ +0.1}_{
 -0.1}$ & $0.065^{ +0.08}_{ -0.07}$ & $0.01^{ +0.02}_{
 -0.00}$ & 628.1/696 & 0.90\\
 & 93/12/24 & $1.57^{ +0.03}_{ -0.03}$ & $0.41^{ +0.08}_{
 -0.09}$ & $0.00^{+0.06}_{ -0.00}$ & $0.000^{  +0.004}_{
 -0.000}$ & 634.4/660 & 0.96\\
\hline
NGC3516 & 94/04/02 & $1.71^{ +0.01}_{ -0.01}$ & $0.74^{ +0.03}_{
 -0.03}$ & $0.39^{ +0.02 }_{-0.03}$ & $0.0^{+0.0}_{
 -0.0}$ & 924.3/772 & 1.20\\
 & 95/03/11 & $1.72^{ +0.03}_{ -0.02}$ & $0.64^{ +0.05}_{ -0.05}$ &
 $0.21^{ +0.05}_{ -0.03}$ & $0.05^{ +0.01}_{ -0.01}$ &
 1025.6/1005 & 1.02\\
 & 95/03/12 & $1.78^{ +0.01}_{ -0.02}$ & $0.64^{ +0.04}_{ -0.04}$ &
 $0.16^{ +0.04}_{ -0.04}$ & $0.028^{ +0.005}_{ -0.007}$ &
 1079.1/1002 & 1.08\\
\hline
NGC4051 & 93/07/27 & $2.37^{ +0.01}_{ -0.01}$ & $0.01^{ +0.03}_{
 -0.01}$ & $0.55^{ +0.03}_{ -0.03}$ & $0.0^{ +0.0}_{
 -0.0}$ & 1320.4/748 & 1.77\\
 & 96/06/27 & $2.219^{ +0.010}_{ -0.004}$ & $0.00^{  +0.01}_{
 -0.00}$ & $0.25^{ +0.01}_{ -0.01}$ & $0.0^{ +0.0}_{
 -0.0}$ &  1746.6/1010 & 1.73\\
\hline
NGC5548 & 93/07/27 & $1.92^{ +0.01}_{ -0.01}$ & $0.28^{
 +0.03}_{ -0.03}$ & $0.14^{ +0.03}_{  -0.03}$  & $0.0^{
 +0.0}_{ -0.0}$ &  829.8/772 & 1.07\\
 & 96/06/27 & $1.91^{ +0.02}_{ -0.02}$ & $0.3631^{ +0.04}_{
 -0.04}$ & $0.08^{ +0.03}_{  -0.03}$  & $0.047^{ +0.005}_{
 -0.005}$ & 1156.1/1011 & 1.14\\
 & 96/06/29 & $1.89^{ +0.02}_{ -0.01}$ & $0.65^{ +0.04}_{
 -0.04}$ & $0.000^{+0.009}_{ -0.000}$  & $0.031^{ +0.006}_{
 -0.005}$ & 1249.5/999  & 1.25\\
 & 96/07/01 & $1.84^{ +0.02}_{ -0.02}$ & $0.69^{ +0.05}_{
 -0.06}$ & $0.016^{ +0.04}_{ -0.02}$  & $0.041^{ +0.007}_{
 -0.007}$ & 1096.4/1002 & 1.09\\
 & 96/07/03 & $1.75^{ +0.03}_{ -0.02}$ & $0.56^{ +0.04}_{
 -0.06}$ & $0.01^{ +0.05}_{ -0.01}$ & $0.03^{ +0.01}_{
 -0.01}$ & 995.5/984  & 1.01\\
 & 96/07/04 & $1.77^{ +0.02}_{ -0.01}$ & $0.56^{+0.02}_{
 -0.05}$ & $0.00^{ +0.04}_{-0.00}$ & $0.03^{ +0.01}_{
 -0.00}$ & 1105.1/1008 & 1.10\\
 & 98/06/15 & $1.94^{ +0.01}_{ -0.01}$ & $0.28^{ +0.02}_{
 -0.03}$ & $0.00^{+0.02}_{ -0.00}$  & $0.046^{ +0.004}_{
 -0.004}$ & 988.4/774  & 1.28\\
 & 98/06/20 & $1.944^{ +0.004}_{ -0.004}$ & $0.281^{+0.009}_{
 -0.015}$ & $0.000^{  +0.006}_{ -0.000}$  & $0.051^{ +0.001}_{
 -0.001}$ & 2115.6/774  & 2.73\\
 & 98/07/07 & $1.78^{ +0.02}_{ -0.01}$ & $0.16^{ +0.03}_{
 -0.05}$ & $0.01^{ +0.04}_{  -0.01}$  & $0.03^{ +0.01}_{
 -0.01}$ & 1006.4/902  & 1.12\\
 & 99/01/19 & $1.84^{ +0.03}_{ -0.03}$ & $0.24^{ +0.07}_{
 -0.06}$ & $0.03^{ +0.05}_{ -0.03}$  & $0.06^{ +0.01}_{
 -0.01}$ & 697.0/664  & 1.05\\
\hline
\medskip
\end{tabular}
\end{center}
\caption{Spectral fitting absorption results for the phenomenological
  model, as defined in Section 3.2 of the main text. Column 3 shows
  the best fitting power law photon index. Column 4 and 5 give the
  optical depths at threshold of the OVII and OVIII absorption edges
  respectively.  Column 6 reports the best fitting column density of
  intrinsic neutral absorbing material (placed at the redshift of the
  source). Column 7 gives the goodness of the fit parameter and the
  number of degrees of freedom (dof). Finally, column 8 presents the
  Red. $\chi^2$ of the fit.}
\end{minipage}
\end{table*}

\subsubsection{Soft excess}
A detailed study of the soft excess was not performed since soft
excesses are expected to be most noticeable below $\sim0.6$ \keV. This
is the energy below which significant SIS calibration uncertainties
exist. However, for completeness and following the analysis in
Reynolds (1997)\nocite{reyno97}, a black-body component (subject to
neutral and ionized absorption) was added to the above model and the
data were re-fitted. To avoid being severely biased by calibration
uncertainties below 0.5 \keV, all data below this energy were
discarded. The results found are presented in Table 3, where a
significant improvement of the fit is obtained for MCG$-$6-30-15, NGC
3516, NGC 4051, and NGC 5548.
\begin{table*}
\begin{minipage}{170mm}
\medskip
\begin{center}
\begin{tabular}{lccccccccc} \hline\hline 
   MCG$-$6-30-15   &&&&&&&&&\\
\hline
Date of     & Photon index & kT    &             &             &    Intrinsic $N_{\rm H}$    & $\chi^2$/dof & Red. $\chi^2$ & $L_{\rm B}$ ($10^{43}$& $L_{\rm B}/L_{\rm 2-10}$\\
Observation & $\Gamma$     & (\keV)& $\tau_{O7}$ & $\tau_{O8}$ &($10^{22}$ $cm^{-2}$) &  & & \ergps) &\\
\hline
1993/07/09 & $1.96^{ +0.02}_{ -0.02}$ & $0.09^{ +0.01}_{ -0.01}$ & $0.61^{ +0.05}_{ -0.04}$ & $0.10^{ +0.03}_{ -0.03}$ & $0.03^{ +0.03}_{ -0.02}$ & 797.5/767  & 1.04 & 0.6 & 0.46\\
1993/07/31 & $1.83^{ +0.02}_{ -0.04}$ & $0.11^{ +0.01}_{ -0.01}$ & $0.8^{ +0.1}_{ -0.1}$ & $0.19^{ +0.04}_{ -0.05}$ & $0.11^{ +0.03}_{ -0.05}$ & 785.8/764  & 1.03 & 1.1  & 1.09\\
1994/07/23 & $1.91^{ +0.01}_{ -0.01}$ & $0.10^{ +0.01}_{ -0.00}$ & $0.65^{ +0.03}_{ -0.01}$ & $0.11^{ +0.01}_{ -0.01}$ & $0.05^{ +0.02}_{ -0.01}$ & 1286.3/1005 & 1.28 & 0.3 & 0.24\\
1997/08/03 & $1.85^{ +0.01}_{ -0.01}$ & $0.080^{+0.004}_{-0.004}$ & $0.78^{ +0.02}_{ -0.02}$ & $0.04^{ +0.01}_{ -0.01}$ & $0.11^{ +0.01}_{ -0.01}$ & 1208.7/1005 & 1.20 & 1.2  & 1.08\\
1997/08/07 & $1.81^{ +0.01}_{ -0.02}$ & $0.10^{ +0.02}_{ -0.01}$ & $0.70^{ +0.04}_{ -0.04}$ & $0.13^{ +0.03}_{ -0.03}$ & $0.10^{ +0.01}_{ -0.03}$ & 1004.6/1007 & 1.00 & 0.5 & 0.46\\
\hline\hline
NGC3516   && &&&&&&&\\
\hline
1994/04/02 & $1.68^{+0.01}_{-0.01}$ & $0.10^{+0.01}_{-0.01}$ & $0.74^{+0.04}_{-0.04}$ & $0.31^{+0.03}_{-0.03}$ & $0.02^{+0.02}_{-0.02}$ & 886.8/767 &  1.16 &  0.3 &  0.1\\
1995/03/11 & $1.73^{+0.03}_{-0.03}$ & $0.14^{+0.03}_{-0.01}$ & $0.66^{+0.05}_{-0.05}$ & $0.25^{+0.04}_{-0.04}$ & $0.05^{+0.02}_{-0.01}$&  1001.2/1000 & 1.0 &0.04 &  0.02\\
1995/03/12 & $1.81^{+0.02}_{-0.03}$ & $0.08^{+0.01}_{-0.02}$ & $0.57^{+0.04}_{-0.07}$ & $0.16^{+0.05}_{-0.04}$ & $0.06^{+0.02}_{-0.02}$ & 1043.1/997 & 1.04 & 0.6 & 0.4\\
\hline\hline
NGC4051  &&  &&&&&&&\\
\hline
1993/04/25 & $2.06^{+0.03}_{-0.02}$ & $0.11^{+0.00}_{-0.01}$ &
 $0.23^{+0.05}_{-0.09}$ &  $0.15^{+0.04}_{-0.06}$ &
 $0.00^{+0.03}_{-0.00}$ &  883.6/743 &  1.19 &  0.6 &  27.1\\
1994/06/07 & $2.01^{+0.02}_{-0.01}$ & $0.13^{+0.01}_{-0.01}$ & $0.22^{
 +0.04}_{-0.03}$ &  $0.12^{+0.02}_{-0.02}$ & $0.02^{+0.03}_{-0.01}$ &
 1145.4/1005 & 1.14 &  0.5 &  15.0\\
\hline\hline
NGC5548   &&&&&&&&&\\
\hline
1993/07/27& $1.86^{+0.02}_{-0.02}$ & $0.16^{+0.01}_{-0.02}$ & $0.37^{+0.03}_{-0.04}$ & $0.15^{+0.03}_{-0.03}$ & $0.00^{+0.01}_{-0.00}$ & 780.0/767 & 1.01 &
 0.4 & 0.1\\
1996/06/27& $1.91^{+0.02}_{-0.02}$ & $0.14^{+0.01}_{-0.02}$ & $0.38^{+0.04}_{-0.04}$ & $0.15^{+0.03}_{-0.03}$ & $0.06^{+0.01}_{-0.01}$ & 1075.2/1006 &  1.07 &  0.5 & 0.1\\
1996/06/29& $1.85^{+0.03}_{-0.03}$ & $0.16^{+0.02}_{-0.03}$ & $0.73^{+0.04}_{-0.05}$ & $0.00^{+0.04}_{-0.00}$ & $0.03^{+0.01}_{-0.01}$ & 1164.9/994 & 1.17 & 0.5 & 0.1\\
1996/07/01& $1.82^{+0.04}_{-0.03}$ & $0.16^{+0.02}_{-0.02}$ &
 $0.70^{+0.06}_{-0.06}$ & $0.10^{+0.05}_{-0.04}$ &
 $0.05^{+0.01}_{-0.01}$ & 1028.2/997 & 1.03 & 0.4 & 0.1\\
1996/07/03& $1.80^{+0.02}_{-0.02}$ & $0.06^{+0.01}_{-0.00}$ & $0.3^{ +0.1}_{-0.1}$ & $0.00^{+0.03}_{-0.00}$ & $0.12^{+0.03}_{-0.02}$ & 915.7/979 & 0.94 & 25.9 & 5.1\\
1996/07/04& $1.78^{+0.03}_{-0.02}$ & $0.14^{+0.02}_{-0.02}$ & $0.6^{+0.1}_{-0.1}$& $0.07^{+0.04}_{-0.04}$ & $0.05^{+0.01}_{-0.01}$ & 1021.1/1003 & 1.02 & 0.8 & 0.1\\
1998/06/15& $1.96^{+0.02}_{-0.02}$ & $0.11^{+0.01}_{-0.01}$ & $0.32^{+0.03}_{-0.03}$ & $0.04^{+0.03}_{-0.03}$ & $0.07^{+0.02}_{-0.01}$ & 828.8/769 & 1.10 & 1.7 & 0.2\\
1998/06/20& $1.96^{+0.01}_{-0.01}$& $0.132^{+0.003}_{-0.004}$ & $0.34^{+0.01}_{-0.02}$ & $0.07^{+0.02}_{-0.01}$ & $0.08^{+0.00}_{-0.01}$ & 1028.7/769 & 1.34 & 1.8 & 0.2\\
1998/07/07& $1.84^{+0.03}_{-0.02}$ & $0.15^{+0.01}_{-0.01}$ & $0.23^{+0.06}_{-0.05}$ & $0.12^{+0.05}_{-0.04}$ & $0.07^{+0.02}_{-0.01}$ & 772.8/740 & 1.04 & 0.9 & 0.2\\
1999/01/19& $1.81^{+0.03}_{-0.06}$ & $0.19^{+0.05}_{-0.03}$ & $0.25^{+0.08}_{-0.05}$ & $0.10^{+0.06}_{-0.06}$ & $0.07^{+0.01}_{-0.01}$ & 674.6/659 & 1.02 & 0.4 & 0.1\\
\hline
\medskip
\end{tabular}
\end{center}
\caption{Spectral fitting results for the phenomenological model
  including a black body component for MCG$-$6-30-15, NGC 3516, NGC
  4051 and NGC 5548. Column 2 gives the best fitting power law photon
  index. Column 3 presents the best fitting black body temperature.
  Column 4 and 5 give the optical depths at threshold of the OVII and
  OVIII absorption edges respectively.  Column 6 reports the best
  fitting column density of intrinsic neutral absorbing material
  (placed at the redshift of the source). Column 7 gives the goodness
  of the fit parameter and the number of degrees of freedom (dof).
  Column 8 presents the Red.  $\chi^2$ of the fit. Column 9 reports
  the total (bolometric) luminosity of the best fitting black body and
  Column 10 shows the ratio of this bolometric luminosity to the 2--10
  \keV luminosity.}
\end{minipage}
\end{table*}
\begin{table*}
\begin{minipage}{170mm}
\medskip
\begin{center}
\begin{tabular}{lccccccc} \hline\hline
Source & Date of & $\log N_{\rm wa}$  & $\xi$ & Intrinsic $N_{\rm H}$ &$L_{\rm 2-10}$& $\chi^2$/dof & Red.\\
& Observation & \pcmsq & (\ergpcmsqps)& ($10^{22}$ \pcmsq)&($10^{43}$ \ergps)& & $\chi^2$\\
\hline
IC4329a &1993/08/15 & $21.49^{+0.04}_{-0.04}$ & $1.4^{+0.5}_{-0.4}$ & $0.24^{+0.01}_{-0.01}$ &7.4&1091.3/773  &1.41\\
&1997/08/07 &  $21.1^{+0.1}_{-0.1}$ & $1.0^{+0.9}_{-0.3}$ &$0.39^{+0.0}_{-0.2}$ &10.1 & 1314.6/1008 & 1.30\\
&1997/08/10 & $21.3^{+0.1}_{-0.1}$ & $1.6^{+0.4}_{-0.8}$ &$0.31^{+0.02}_{-0.01}$ & 7.5 & 824.6/775 & 1.07\\
&1997/08/12 &  $21.2^{+0.1}_{-0.1}$ & $4.^{+3.}_{-3.}$ &$0.42^{+0.02}_{-0.02}$ & 8.6 & 899.2/773 & 1.16\\
&1997/08/15 & $21.79^{+0.02}_{-0.03}$ & $0.3^{+0.5}_{-08.}$&$0.03^{+0.03}_{-0.02}$ & 10.4 & 1283.5/1006 & 1.28\\
\hline
MCG$-$6-30-15 &1993/07/09 & $21.77^{+0.02}_{-0.02}$ & $8.2^{+0.3}_{-0.3}$ & $0.^{+0.}_{ -0.}$ &1.3&1159.3/773 & 1.50\\
&1993/07/31 &$21.96^{+0.03}_{-0.04}$ & $12.^{+1.}_{-1.}$  &  $0.^{+0.}_{-0.}$ & 1.1&1115.5/770 & 1.45\\
&1994/07/23 &$21.62^{+0.04}_{-0.01}$ & $4.4^{+0.6}_{-0.6}$ &$0.^{+0.}_{-0.}$ & 1.2& 1.933.2/1011 & 1.91\\
&1997/08/03 & $21.70^{+0.08}_{-0.02}$ & $4.2^{+0.1}_{-0.1}$ & $0.^{+0.}_{-0.}$ &1.0& 1568.5/1010 & 1.55\\
&1997/08/07 &$21.67^{+0.01}_{-0.01}$ & $4.1^{+0.1}_{-0.1}$ &  $0.^{+0.}_{-0.}$ &0.9& 1312.3/1013 & 1.30\\
\hline
MR 2251-178 &1993/11/16 & $21.6^{+0.1}_{-0.1}$ &$9.^{+2.}_{-1.}$ &$0.^{+0.}_{-0.}$ & 81.4&771.5/759 & 1.02\\
&1993/12/07 & $21.5^{+0.1}_{-0.1}$ & $8.^{+1.}_{-1.}$ & $0.^{+0.}_{-0.}$ &91.4 & 840.3/868 & 0.97\\
&1993/12/14 &$21.5^{+0.1}_{-0.1}$ & $8.^{+2.}_{-2.}$  & $0.^{+0.}_{-0.}$  &80.8&  776.0/751 & 1.03\\
&1993/12/19 & $ 21.5^{+0.1}_{-0.1}$ &$7.^{+1.}_{-2.}$ & $0.^{+0.}_{-0.}$ &77.3& 635./697 & 0.91\\
&1993/12/24 & $21.4^{+0.1}_{-0.1}$ & $9.^{+4.}_{-1.}$  & $0.^{+0.}_{-0.}$ &62.5& 640.3/661 & 0.97\\
\hline
NGC3516 &1994/04/02 &$21.80^{+0.01}_{-0.01}$ & $9.3^{+0.1}_{-0.0}$ &  $0.^{+0.}_{-0.}$ &2.5& 1475.2/773 & 1.91\\
&1995/03/11 & $21.63^{+0.02}_{-0.02}$  &$3.6^{+0.4}_{-0.3}$ & $0.^{+0.}_{-0.}$ &1.6& 1105.2/1006 & 1.10\\
&1995/03/12 &$21.64^{+0.02}_{-0.02}$ & $4.2^{+0.4}_{-0.3}$ & $0.^{+0.}_{-0.}$ &1.5&  1151.4/1003 & 1.15\\
\hline
NGC4051 &1993/04/25 &$22.45^{+0.03}_{-0.04}$ & $41.^{+2.}_{-3.}$ &  $0.^{+0.}_{-0.}$ &0.03& 1585.1/749 & 2.11\\ 
&1994/06/07 &$22.43^{+0.01}_{-0.02}$ & $48.^{+1.}_{-1.}$ &  $0.^{+0.}_{-0.}$ &0.04& 1982.0/1011 & 1.96\\
\hline
NGC5548 & 1993/07/27 & $21.5^{+0.1}_{-0.0}$ & $11.^{+3.}_{-1.}$ &$0.^{+0.}_{-0.}$ &4.9& 899.5/773 &  1.16\\
 &1996/06/27 &$21.41^{+0.01}_{-0.03}$ & $3.3^{+0.5}_{-0.7}$ & $0.03^{+0.00}_{-0.00}$ &7.6& 1196.1/1012 & 1.18\\
 &1996/06/29 &$21.40^{+0.02}_{-0.02}$ & $1.6^{+0.2}_{-0.2}$ & $0.^{+0.}_{-0.}$ &6.8& 1288.8/1000 & 1.29\\
 &1996/07/01 &$21.45^{+0.01}_{-0.02}$ & $1.6^{+0.2}_{-0.1}$ & $0.^{+0.}_{-0.}$ &5.6&  1140.6/1003 & 1.14\\
 &1996/07/03 &$21.45^{+0.02}_{-0.02}$ & $1.8^{+0.4}_{-0.3}$ & $0.^{+0.}_{-0.}$ & 4.8& 999.5/985 & 1.01\\
 &1996/07/04 &$21.40^{+0.01}_{-0.03}$ & $1.8^{+0.2}_{-0.3}$ & $0.^{+0.}_{-0.}$ &5.9&  1130.8/1009 & 1.12\\
 &1998/06/15 &$21.19^{+0.03}_{-0.02}$ & $2.0^{+1.0}_{-0.2}$ & $0.04^{+0.00}_{-0.00}$ &7.0&  1018.8/775 & 1.31\\
 &1998/06/20 &$21.16^{+0.03}_{-0.06}$ & $1.1^{+0.4}_{-0.6}$ & $0.03^{+0.00}_{-0.00}$ &8.2& 2447.0/775 & 2.90\\
 &1998/07/07 &$21.15^{+0.02}_{-0.06}$ & $0.4^{+0.3}_{-0.1}$ & $0.^{+0.}_{-0.}$ &5.1& 849.1/746 & 1.14\\
 &1999/01/19 &$21.15^{+0.07}_{-0.05}$ & $4.^{+2.}_{-1.}$ &  $0.05^{+0.01}_{-0.01}$ &5.5& 702.8/665 & 1.06\\
\hline
\medskip
\end{tabular}
\end{center}
\caption{Spectral fitting results using the one-zone photoionization model.
  Columns 3 and 4 give the best-fitting column density and ionization
  parameter $\xi$ of the warm photoionized plasma. Column 5 gives the
  best fitting column density of intrinsic neutral absorbing material
  (placed at the redshift of the source). Column 6 reports the
  goodness of the fit parameter and column 7 the Red.  $\chi^2$ of the
  fit.}
\end{minipage}
\end{table*}

\subsubsection{Photoionization models}
It will be assumed that the physics of the WA is dominated by
photoionization. This assumption results from the very high and
variable ionization state. Quantitative evidence for this comes from
the observed anti-correlation between the OVIII absorption edge depth
and the primary ionizing flux observed during the long observation of
MCG$-$6-30-15 \cite{ota96}. Also NGC 4051 has shown variability of the
OVII edge (Guainazzi et al. 1996). A grid of self-consistent models
has been produced using Ferland's photoionization code Cloudy C90.04
\cite{Ferl98}. Initially, a power-law ionizing continuum has been
used, of photon index $\Gamma$ (determined by the phenomenological
model above), and luminosity between 2--10 \keV, $L_{\rm 2-10}$ equal
to $10^{43}$ \ergps, incident onto a shell at inner radius $r=10^{16}$
\cm of density $n=5\times10^9$--$10^{13}$ \pcm.  Therefore, the
ionization parameter:
\begin{equation}
\xi=\frac{L_{2-10}}{nr^2}
\end{equation}
varies from 0.01 to 20 \ergcmps. The column density was varied from
$N_{\rm wa}=10^{20.3}$--$10^{24}$ \pcmsq. NGC 4051 is the only object
in our sample that has $\Gamma>2$ and also the only source presenting
absorption features at energies larger than 1 \keV. It required larger
values for the ionization parameter: 1--50 \ergcmps; hence the range
of values covered by the density is $n=5\times10^{10}$--$10^{11}$
\pcm. We will be assuming a constant density cloud in our modelling,
which may not be true in every case (e.g. \citeNP{Binette98}). A black
body has been included for MCG$-$6-30-15, NGC 3516, NGC 4051, and NGC
5548 as given in Table 3.  Recall that spectral variability of
MCG$-$6-30-15 strongly argues for a multi-zone warm absorber (e.g.
Otani et al. 1996\nocite{ota96}, Morales, Fabian and Reynolds
2000\nocite{Mora00a}, \shortciteNP{Lee01}). The simultaneous UV/X-ray
studies of NGC3516 by Kriss et al.
(1996a,b)\nocite{Kris96a}\nocite{Kris96b} also imply the existence of
a complex stratified ionized absorber. The \chandra observation of NGC
4051 \shortcite{Coll01} has revealed a two-zone WA for this object.
Thus, one-zone photoionization models can be only regarded as a useful
parameterization of the spectral data.

All the observations for each source were examined using Cloudy models
of the type described above. The results of these fits are shown in
Table 4. The data still show deviations from this best fitting at
$\sim$ 1 \keV.  This presumably signals a breakdown in one of the
assumptions of this model (such as the one-zone nature or the chemical
abundances of the plasma). 

An illustration of the incident and transmitted continuum as
calculated by our models and the confidence contour plot for two
observations of MCG$-$6-30-15 and NGC 5548 is illustrated in Fig.
1\footnote{The incident and transmitted continuum as calculated by our
  models for all the observations as well as all the contour plots are
  presented at \url{http://www-xray.ast.cam.ac.uk/~rm/BH_mass/}}. In
Table 5 the values for $L_{\rm abs}$ for each of the 30 observations
is reported.
\begin{figure*}
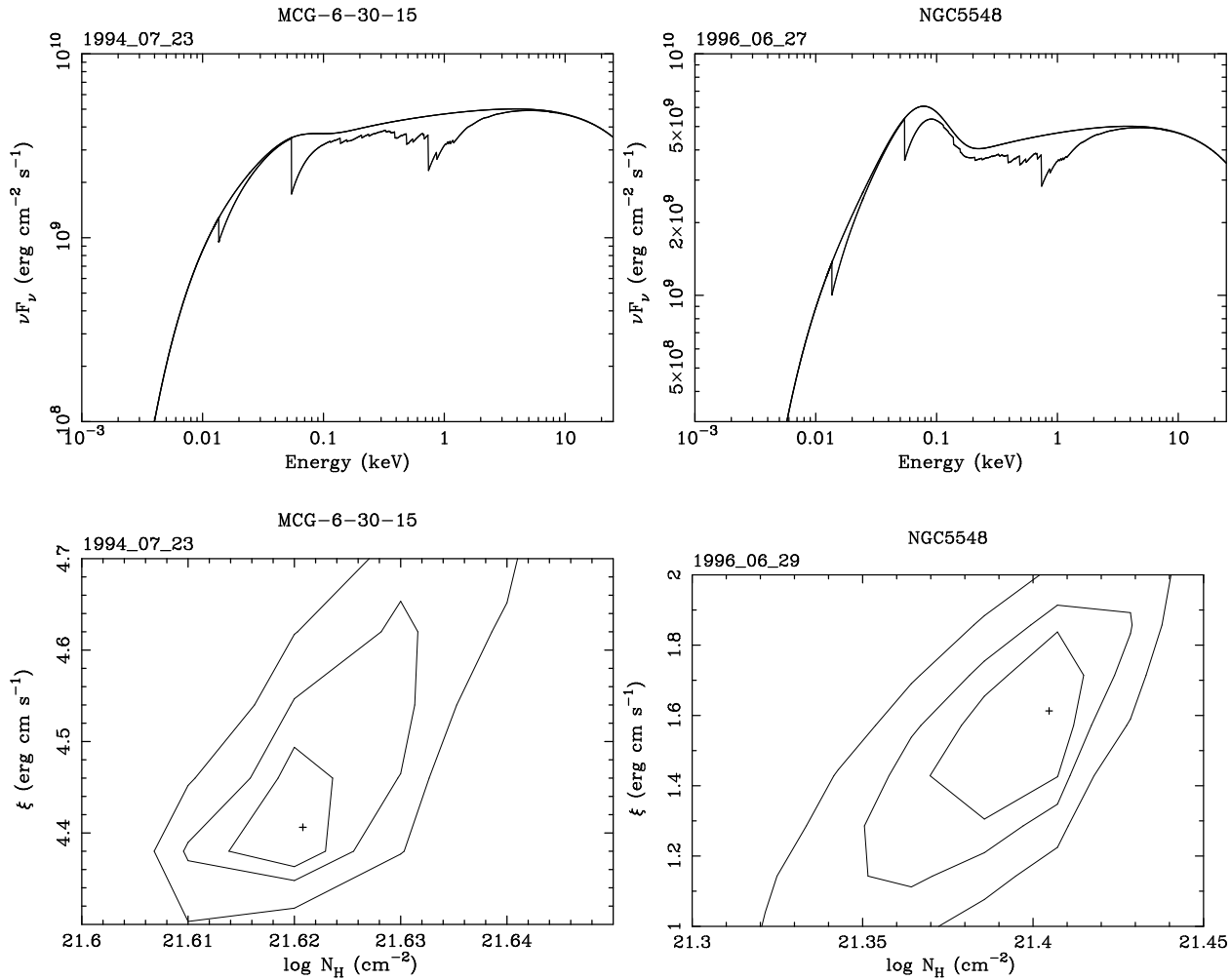

\begin{minipage}{150mm} 
\psrotatefirst 
\centerline{
        \psfig{file=method_MCG-6-30-15_1994_07_23.ps,width=3.25in,angle=-90,silent=}
        \psfig{file=method_NGC5548_1996_06_27.ps,width=3.25in,angle=-90,silent=}
}
\vspace{5mm} 
\centerline{
        \psfig{file=MCG-6-30-15_1994_07_23_contour.ps,width=3.25in,angle=-90,silent=}
        \psfig{file=NGC5548_1996_06_29_contour.ps,width=3.25in,angle=-90,silent=}
}
\caption{Top panels: example of the incident continuum (top line) and
  transmitted continuum (bottom line) as predicted by our models for
  the 1994/07/23 MCG$-$6-30-15 and the 1996/06/27 NGC 5548
  observations. The area enclosed by the two lines gives the value
  for the flux absorbed by the WA clouds. Bottom panels: contour plots
  for the two observations in the plane ionization parameter, $\xi$,
  against column density, $N_{\rm wa}$ at the 68, 90 and 99 per cent
  confident level for the two parameters of interest.}
\end{minipage}
\end{figure*}
\begin{table}
\medskip
\begin{center}
\begin{tabular}{lcc} \hline\hline
Source & Date of & $\log L_{\rm abs}$\\
 & Observation & (\ergps)\\
\hline
IC4329a & 93/08/15 & $42.6\pm0.3$ \\
 & 97/08/07 & $42.3\pm0.3$ \\
 & 97/08/10 & $42.4\pm0.3$ \\
 & 97/08/12 & $42.2\pm0.3$ \\
 & 97/08/15 & $43.0\pm0.3$ \\
\hline
MCG$-$6-30-15 & 93/07/09 & $42.60\pm0.05$ \\
 & 93/07/31 & $42.70\pm0.05$ \\
 & 94/07/23 & $42.59\pm0.05$ \\
 & 97/08/03 & $42.67\pm0.05$ \\
 & 97/08/07 & $42.64\pm0.05$ \\
\hline
MR 2251-178 & 94/04/02 & $42.42\pm0.06$ \\
 & 93/12/07 & $42.37\pm0.06$ \\
 & 93/12/14 & $42.33\pm0.06$ \\
 & 93/12/19 & $42.37\pm0.06$ \\
 & 93/12/24 & $42.24\pm0.06$ \\
\hline
NGC3516 & 94/04/02 & $42.608\pm0.001$ \\
 & 95/03/11 & $42.609\pm0.001$ \\
 & 95/03/12 & $42.606\pm0.001$ \\
\hline
NGC4051 & 93/04/25$^*$ & $42.85\pm0.05$ \\
        & 94/06/07     & $42.80\pm0.05$ \\
\hline
NGC5548 & 93/07/27 & $42.3\pm0.2$ \\
 & 96/06/27 & $42.4\pm0.2$ \\
 & 96/06/29 & $42.5\pm0.2$ \\
 & 96/07/01 & $42.2\pm0.2$ \\
 & 96/07/03 & $42.6\pm0.2$\\
 & 96/07/04 & $42.5\pm0.2$\\
 & 98/06/15 & $42.3\pm0.2$ \\
 & 98/06/20$^*$ & $42.4\pm0.2$\\
 & 98/07/07 & $42.5\pm0.2$\\
 & 99/01/19 & $42.1\pm0.2$\\
\hline
\medskip
\end{tabular}
\end{center}
\caption{Luminosity absorbed by the WA. Column 1 gives the name of the source,
 column 2 indicates the date (year/month/day) of the observation and
 column 3 the values of the luminosity absorbed, $L_{\rm abs}$. Those
 observations marked with an asterisk, *, have a value of $\chi^2>2$
 when the data is  fitted to our photoionization model (see Table 4).}
\end{table} 

\subsection{Results and comparison with other estimates}
\begin{figure*}
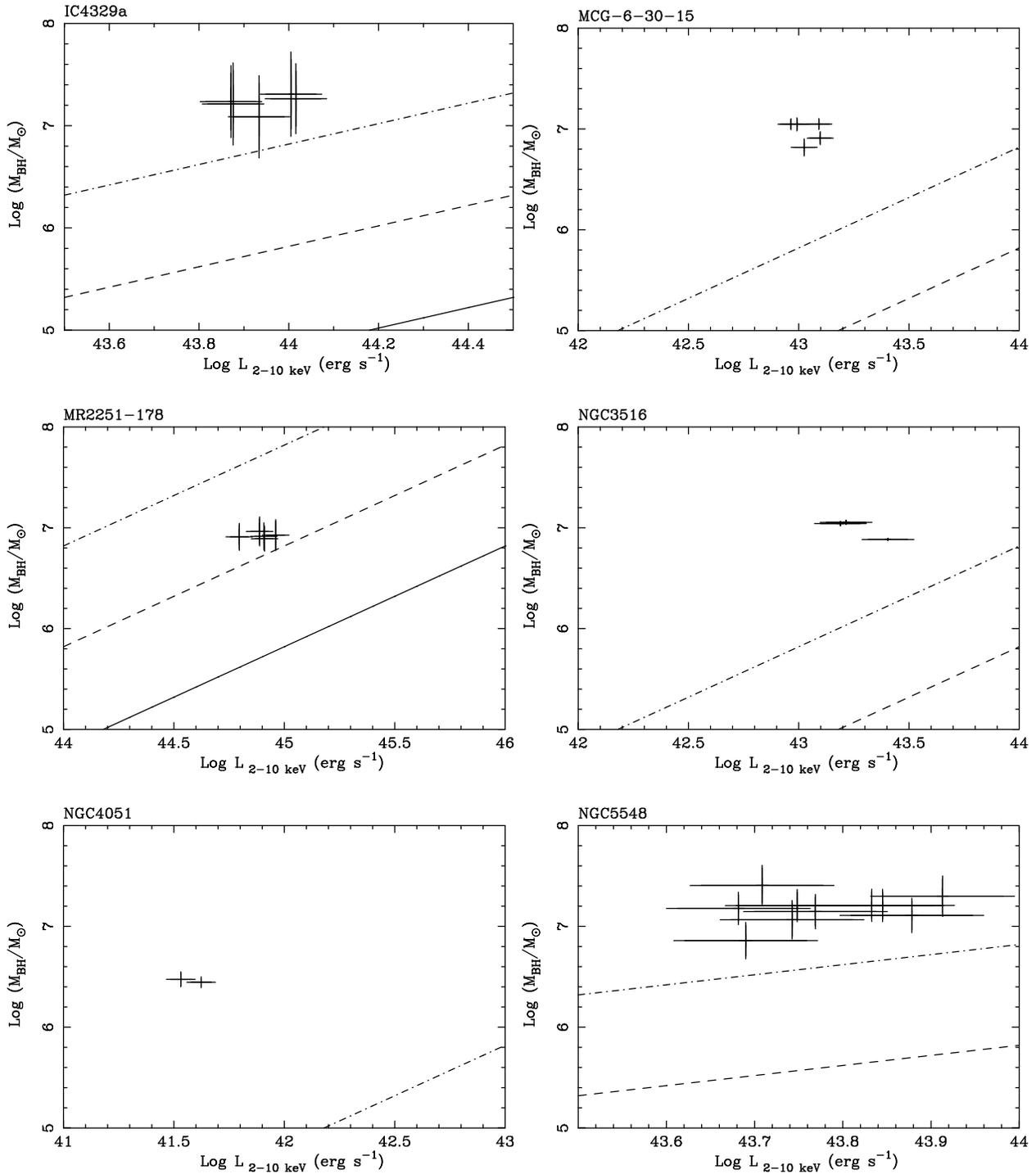

\begin{minipage}{150mm} 
\psrotatefirst 
\centerline{
        \psfig{file=IC4329a_Mbh_vs_L_2_10.ps,width=3.25in,angle=-90,silent=}
        \psfig{file=MCG-6-3_Mbh_vs_L_2_10.ps,width=3.25in,angle=-90,silent=}
}
\vspace{5mm} 
\centerline{
        \psfig{file=MR2251-_Mbh_vs_L_2_10.ps,width=3.25in,angle=-90,silent=}
        \psfig{file=NGC3516_Mbh_vs_L_2_10.ps,width=3.25in,angle=-90,silent=}
}
\vspace{5mm} 
\centerline{
        \psfig{file=NGC4051_Mbh_vs_L_2_10.ps,width=3.25in,angle=-90,silent=}
        \psfig{file=NGC5548_Mbh_vs_L_2_10.ps,width=3.25in,angle=-90,silent=}
}
\caption{Black hole mass, $M_{\rm BH}$, in solar units for the 30 
  observations of the 6 AGN (the names of the objects appear at the
  top left corner of each figure) versus the 2--10 \keV luminosity,
  $L_{\rm 2-10}$.  The solid line corresponds to $L=L_{\rm Edd}$, the dashed
  line to $L=0.1L_{\rm Edd}$, and the dot-dashed line to $L=0.01L_{\rm Edd}$,
  where $L=10L_{\rm 2-10}$.}
\end{minipage}
\end{figure*}

\subsubsection{Mean values, errors and comparison with other methods}
We have substituted the results of $L_{\rm abs}$ and $N_{\rm wa}$ from
each observation in Eqn. 2. The value of the black hole mass, $M_{\rm
  BH}$ for each observation versus the 2--10 \keV luminosity, $L_{\rm
  2-10}$ is presented in Fig. 2 and reported in Table 6. There are
differences in the results for $M_{\rm BH}$ obtained for each
observation that can be accounted for within the errors, except for
one observation of NGC 3516\footnote{See Table 4. The 2--10 luminosity
  for this source in the 1994 observation is approximately 2 times
  larger than in those observations in 1995.  Also note how our
  photoionization model does not reproduce very well this very first
  observation, giving a $\chi ^2=1.91$.}. A possible cause for these
differences could be the dynamical evolution of the WA, since we would
be proving different material for different observations. In
particular for the case of the NGC 3516 UV and X-ray absorber, such
evolution has been reported (\citeNP{Koratkar96} and \citeNP{Kris96a}).

It is worth noting that all the MR2251-178 observations were taken
within one month and for this particular object the consistency of the
results is remarkable. There is also a variation in $L_{\rm 2-10}$
with time for all the sources (i.e. the source gets brighter or
fainter). These changes in the X-ray luminosity do not seem to relate
to the values obtained for $M_{\rm BH}$ in a simple manner.

For several of these AGN, there is extensive observational evidence
for a multi-zone WA. For the case of MCG$-$6-30-15, time variability
of the OVIII edge (e.g. \shortciteNP{ota96}, \shortciteNP{Mora00a}) and
the detection of different ionization species \shortcite{Lee01}
support the multi-zone hypothesis. And very recently, for the case of
NGC 4051, \shortciteN{Coll01} have resolved two distinct absorption
systems. We have not taken into account the multi-zone nature of the WA
in the present work.

Another element that might have some effect on our results is that
below approximately 1 \keV SIS0 and SIS1 spectra since approximately
late 1994 have been showing an increasing divergence from each other,
and from the GIS data, towards lower energies. This is in the sense
that both SIS0 and SIS1 efficiencies below 1 \keV have been steadily
decreasing over time. This loss in efficiency is currently not
corrected by any of the software and a detailed analysis of how this
could affect our results is discussed in Section 3.3.2.
\begin{table}
\medskip
\begin{center}
\begin{tabular}{lccc} \hline\hline
Source & Date of & $\log L_{\rm 2-10}$& $\log M_{\rm BH}/M_{\odot}$\\
 & Observation & (\ergps) & \\
\hline
IC4329a & 93/08/15 & $43.87\pm0.07$ & $7.2^{+ 0.4}_{ -0.4}$ \\
 & 97/08/07 & $44.01\pm0.07$ & $7.3^{+ 0.4}_{ -0.4}$ \\
 & 97/08/10 & $43.88\pm0.07$ & $7.2^{+ 0.4}_{ -0.4}$ \\
 & 97/08/12 & $43.93\pm0.07$ & $7.1^{+ 0.4}_{ -0.4}$ \\
 & 97/08/15 & $44.02\pm0.07$ & $7.3^{+ 0.3}_{ -0.3}$ \\
\hline
MCG$-$6-30-15 & 93/07/09 & $43.10\pm0.06$ & $6.90^{+ 0.07}_{ -0.07}$ \\
 & 93/07/31 & $43.03\pm0.06$ & $6.81^{+0.09}_{ -0.08}$\\
 & 94/07/23 & $43.09\pm0.06$ & $7.05^{+0.06}_{ -0.08}$\\
 & 97/08/03 & $42.99\pm0.06$ & $7.05^{+0.07}_{ -0.13}$\\
 & 97/08/07 & $42.96\pm0.06$ & $7.05^{+0.06}_{ -0.06}$\\
\hline
MR 2251-178 & 94/04/02 & $44.91\pm0.06$ & $6.9^{+0.1}_{ -0.1}$\\
 & 93/12/07 & $44.96\pm0.06$ & $6.9^{+0.1}_{ -0.1}$\\
 & 93/12/14 & $44.90\pm0.06$ & $6.9^{+0.1}_{ -0.1}$\\
 & 93/12/19 & $44.88\pm0.06$ & $7.0^{+0.1}_{ -0.1}$\\
 & 93/12/24 & $44.80\pm0.06$ & $6.9^{+0.1}_{ -0.1}$\\
\hline
NGC3516 & 94/04/02 & $43.4\pm0.1$ & $6.88^{+0.01}_{ -0.01}$\\
 & 95/03/11 & $43.2\pm0.1$ & $7.05^{+0.02}_{ -0.02}$\\
 & 95/03/12 & $43.2\pm0.1$ & $7.04^{+0.02}_{ -0.02}$\\
\hline
NGC4051 & 93/04/25$^*$ & $41.53\pm0.07$ & $6.47^{+0.08}_{ -0.07}$\\
        & 94/06/07 & $41.62\pm0.07$ & $6.45^{+0.06}_{ -0.05}$\\
\hline
NGC5548 & 93/07/27 & $43.69\pm0.07$ & $6.9^{+0.2}_{ -0.2}$\\
 & 96/06/27 & $43.87\pm0.07$ & $7.1^{+0.2}_{ -0.2}$\\
 & 96/06/29 & $43.83\pm0.07$ & $7.2^{+0.2}_{ -0.2}$\\
 & 96/07/01 & $43.75\pm0.07$ & $7.2^{+0.2}_{ -0.2}$\\
 & 96/07/03 & $43.68\pm0.07$ & $7.2^{+0.2}_{ -0.2}$\\
 & 96/07/04 & $43.77\pm0.07$ & $7.2^{+0.2}_{ -0.2}$\\
 & 98/06/15 & $43.85\pm0.07$ & $7.2^{+0.2}_{ -0.2}$\\
 & 98/06/20$^*$ & $43.91\pm0.07$ & $7.3^{+0.2}_{ -0.2}$\\
 & 98/07/07 & $43.70\pm0.07$ & $7.4^{+0.2}_{ -0.2}$\\
 & 99/01/19 & $43.74\pm0.07$ & $7.1^{+0.2}_{ -0.2}$\\
\hline
\medskip
\end{tabular}
\end{center}
\caption{Black hole mass, $M_{\rm BH}$, in solar units for the 30
 observations of our study. Column 1 gives the name of the source,
 column 2 indicates the date (year/month/day) of the observation and
 columns 3 and 4 give the 2 to 10 \keV luminosity, $L_{\rm 2-10}$, and the
 black hole mass, $M_{\rm BH}$, in solar units. Those observations marked
 with an asterisk, *, have a value of $\chi^2>2$ when the data is 
 fitted to our photoionization model (see Table 4).}
\end{table} 

Once the value of the black hole mass was calculated for each
observation, we have obtained the mean value for each source. For
that, those observations with $\chi^2>2$ have been excluded. Our
result for NGC 4051 might not be significant giving the high value of
$\chi^2$ for both observations (although our result is consistent with
other methods estimates). The errors only take into account the
uncertainties in $L_{\rm abs}$ (calculated from the dispersion in the
value of $L_{\rm abs}$ for each observation) and $N_{\rm wa}$
(obtained from the fits). The mean values for $M_{\rm BH}$ are
presented in Figure 3 together with other independent methods
estimates: reverberation mapping \shortcite{Wand99} for IC4329a, NGC
4051, and NGC 5548; X-ray variability \shortcite{Czer00} for
MCG$-$6-30-15, NGC 4051, and NGC 5548; and velocity dispersion results
\shortcite{Gebh00} for NGC 3516 and NGC 4051.  Our estimates for all
sources cluster around $10^7$ $M_{\odot}$. Despite the simplicity of
our model, our results reproduce within an order of magnitude other
independent estimates.
\begin{figure*}
\begin{minipage}{150mm}
\psrotatefirst
\centerline{\psfig{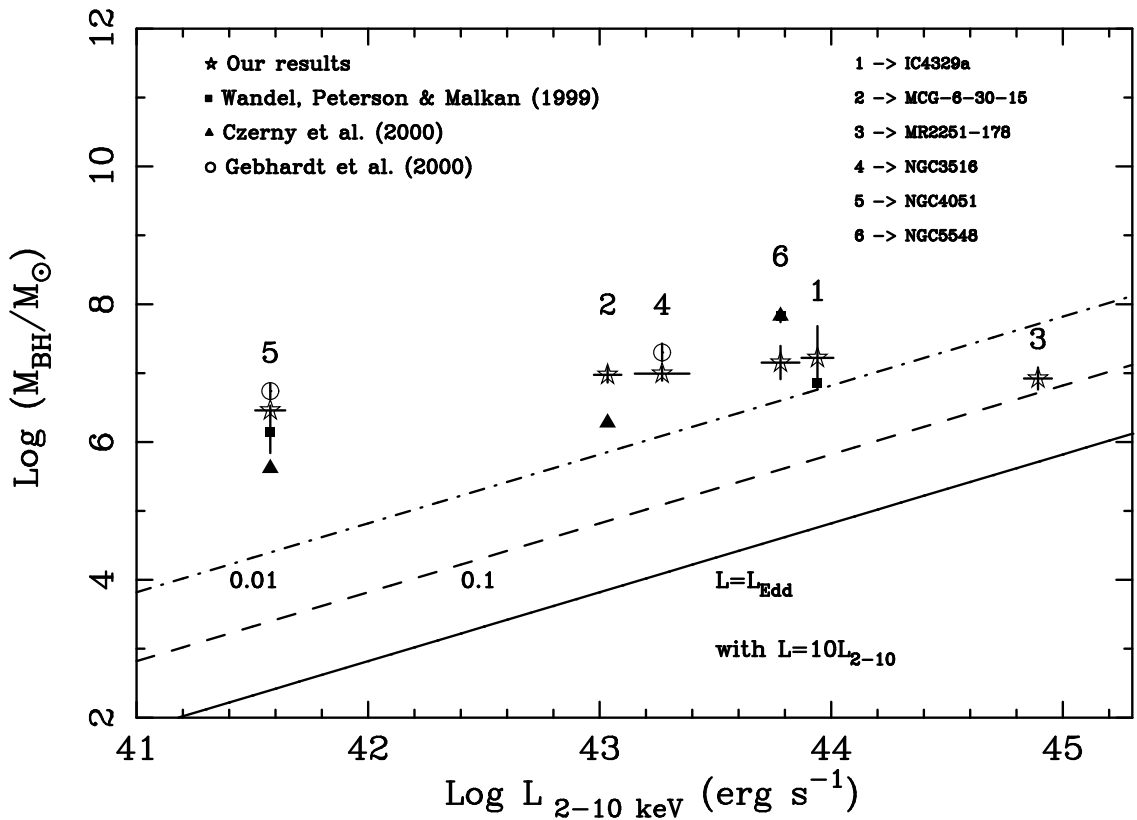}}
\caption{Our results for the mean black hole mass (stars) of the six
  sources versus the mean 2 to 10 \keV luminosity, $L_{\rm 2-10}$.
  Other estimates (reverberation mapping (Wandel, Peterson \& Malkan
  1999), marked as squares for IC4329a, NGC 4051, and NGC 5548, X-ray
  variability (Czerny et al. 1999) , marked as triangles for
  MCG$-$6-30-15, NGC 4051, and NGC 5548, and velocity dispersion
  (Gebhardt et al. 2000), marked as circles for NGC 3516 and NGC 4051)
  have been plotted at our mean $L_{\rm 2-10}$. Reminder: \textbf{our results
  are upper limits on the black hole mass} and are not marked
  as such to avoid confusion. The solid line corresponds to $L=L_{\rm
    Edd}$, the dashed line to $L=0.1L_{\rm Edd}$, and the dot-dashed
  line to $L=0.01L_{\rm Edd}$, where $L=10L_{\rm 2-10}$.}
\end{minipage}
\end{figure*}

\subsubsection{The case of NGC4051}
At low energies, the X-ray continuum of NGC 4051 has been observed to
be dominated by a variable soft excess (e.g.
\shortciteNP{Turn89,Guai96}). \shortciteN{Guai96} also argued for
significant variability of the OVII edge while the OVIII edge remained
roughly constant in strength, precisely the reverse of the behavior
seen in MCG$-$6-30-15 \cite{reyno97}. This object has very recently
been observed with \chandra \cite{Coll01}, and they clearly detect a
two-zone WA.

There are only two observations for NGC 4051 and both of them, when
fitted to our photoionization models, give a $\chi ^2\sim2$. This
indicates that our model breaks down for this particular object.  In
Section 3.2 it was already noted that NGC 4051 was the only object in
our sample that has $\Gamma>2$ and also the only source presenting
absorption features at energies larger than 1 \keV.

We have investigated if the reason for this disagreement could be the
divergence between SIS0 and SIS1 spectra and from the GIS data, since
approximately late 1994. To test this hypothesis we have analysed each
detector individually. Table 7 reports the results and Fig. 4 gives a
graphical representation of them. Surprisingly the only result that is
inconsistent with the rest is one given by GIS2 in the first
observation. Apart from this single value, the other estimates are
consistent with those obtained for the case of the full analysis,
favoring the singular nature of NGC 4051 as the cause for not fitting
a single ionization parameter model.
\begin{table}
\medskip
\begin{center}
\begin{tabular}{lccc} \hline\hline
Date of & \asca&$\log L_{\rm 2-10}$& $\log M_{\rm BH}/M_{\odot}$\\
Observation &apparatus& (\ergps) & \\
\hline
93/04/25 & SIS0&$41.6\pm0.1$ & $6.4^{+ 0.4}_{ -0.4}$ \\
& SIS1& $41.3\pm0.1$ & $6.5^{+ 0.4}_{ -0.4}$ \\
& GIS2& $41.6\pm0.1$ & $7.5^{+ 0.5}_{ -0.5}$ \\
& GIS3& $41.7\pm0.1$ & $6.5^{+ 1.1}_{ -0.4}$ \\
\hline 
94/06/07 & SIS0&$41.6\pm0.1$ & $6.5^{+ 0.4}_{ -0.3}$ \\
& SIS1& $41.7\pm0.1$ & $6.5^{+ 0.3}_{ -0.3}$ \\
& GIS2& $41.6\pm0.1$ & $6.4^{+ 0.5}_{ -0.4}$ \\
& GIS3& $41.6\pm0.1$ & $6.4^{+ 0.4}_{ -0.4}$ \\
\hline
\medskip
\end{tabular}
\end{center}
\caption{Absorbed luminosity and BH mass for NGC 4051. First
  column reports the year/month/day of the observation, second column
 gives the \asca when the 4 \asca instrument used in the fit. Finally
 columns 3 and 4 give the 2--10\keV luminosity, $L_{\rm 2-10}$, and the
 black hole mass, $M_{\rm BH}$, in solar units.}
\end{table}
\begin{figure}
\centerline{\psfig{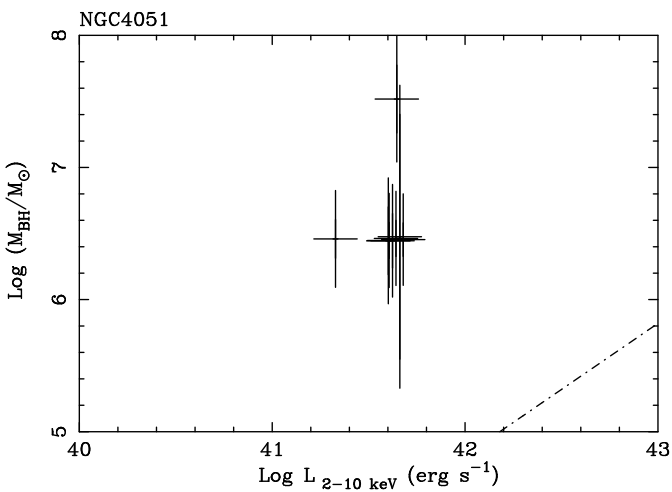}}
\caption{Values obtained for the BH mass of NGC 4051 when the 4
  different \asca instrument are independently fitted for the 1993 and
  1994 observations of our sample. The dot-dashed line corresponds to
  $L=0.01L_{\rm Edd}$, where $L=10L_{\rm 2-10}$.}
\end{figure}

\section{THE INFLUENCE OF THE UNDERLYING CONTINUUM ON OUR ESTIMATE}
The emitted spectrum of Type 1 galaxies in the extreme UV/very soft
X-ray band is poorly known although the photons are highly effective
at ionizing warm absorbing gas. Study of the effect of the excess soft
X-ray emission on our results is therefore of great importance. The
object we have selected to study the influence of a different
underlying continuum has been NGC 5548. The reason for that has been
the availability of a model for its intrinsic spectrum (i.e. after the
effect of Galactic absorption at the soft excess and reddening at the
optical/UV spectrum has been removed), as given by \citeN{Magd98}.
Also for this particular object there are 10 different observation in
our sample, which allow us a better comparison with our results.

Fig. 5, left panel, presents a montage of the incident and transmitted
continuum as calculated by our photoionization models for 4
observations of this object. The different soft excess for each
observation generated a different level of ionization in the WA and
therefore the spectrum for different observations present different
absorption edges (e.g. for observation 1998/07/07, labeled 9 in Fig.
5, the absorption edges for H$_{\rm I}$ and He$_{\rm II}$ at at 0.0136
\keV and 0.0544 \keV, respectively, are very prominent). However,
since the soft X-ray spectrum has to be reproduced by the models after
the ionizing radiation has passed through the WA, the black hole mass
estimate through the ratio $L_{\rm abs}/N_{\rm wa}$ would regulate
itself to changes in either quantity (i.e.  higher values for the
luminosity absorbed demand higher values for the column density and a
strong EUV excess means that He is more highly ionized). As
seen in Table 6, the errors can account for the differences in the
results.

The comparison between the BH mass estimates obtained using our
underlying continuum, $(M_{\rm BH}/M_{\odot})_{\rm cont. a}$, and those
using the \citeN{Magd98} spectrum, $(M_{\rm BH}/M_{\odot})_{\rm cont. b}$,
is illustrated in Fig. 5, right panel. The former results appear to be
lower than the latter, but still the differences are accounted for
within the errors.
\begin{figure*}
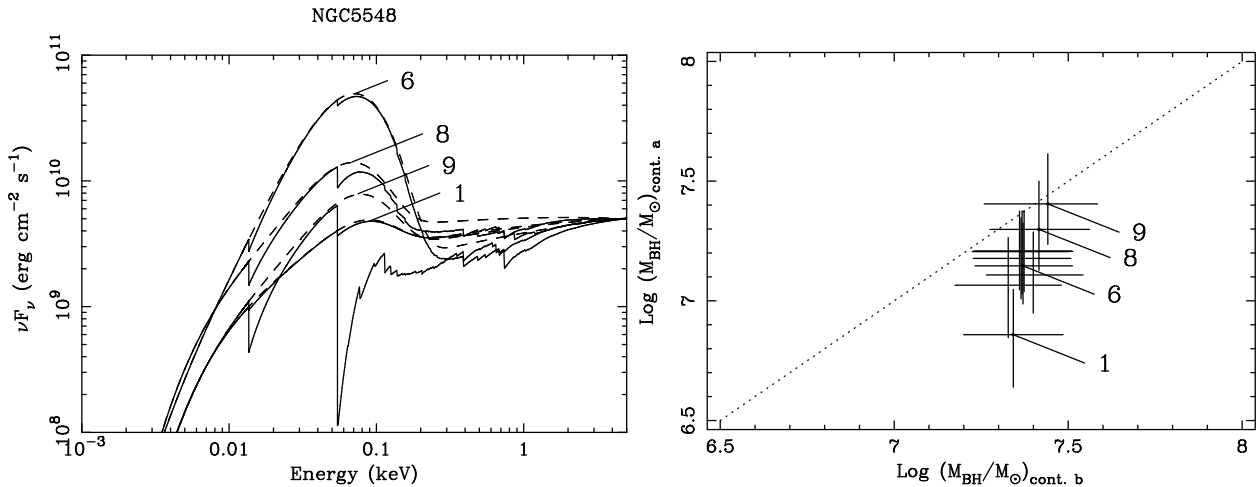

\begin{minipage}{150mm} 
\psrotatefirst 
\centerline{
        \psfig{file=comparison_inc_trans_ngc5548.ps,width=3.25in,angle=-90,silent=}
        \psfig{file=massvsmass.ps,width=3.25in,angle=-90,silent=}
}
\caption{Left panel: Incident (dashed line) and transmitted (solid
  line) continuum for the 1 (1993/07/27), 6 (1996/07/04), 8
  (1998/06/20), and 9 (1998/07/07) observations of NGC 5548, as
  calculated by our photoionization models. This figure shows the
  differences in the incident continuum for these observations, as
  well as graphically presenting the different amounts of luminosity
  absorbed by the WA.  The absorption edges for H$_{\rm I}$ and
  He$_{\rm II}$ at at 0.0136 \keV and 0.0544 \keV, respectively,
  become very prominent for some observations. Right panel: Comparison
  of the results for the BH masses using our underlying continuum,
  $(M_{\rm BH}/M_{\odot})_{\rm cont. a}$, versus those obtained using
  Magdziarz et al. (1998) continuum, $(M_{\rm BH}/M_{\odot})_{\rm
    cont. b}$. The dotted line marks the equality of both values. The
  1 (1993/07/27), 6 (1996/07/04), 8 (1998/06/20), and 9 (1998/07/07)
  observations are labeled.}
\end{minipage}
\end{figure*}

\section{ANALYSIS OF FURTHER TERMS IN THE EQUATION OF MOTION}

\subsection{The drag term}
This section discusses the possibility of a drag force acting on the
WA clouds, and also presents some problems that the cloud model
raises.  For the purposes of estimating the black hole mass, the gas
that is responsible for the WA is assumed to exist as isolated clouds
which are likely moving through an intercloud (IC) medium. These
clouds could be confined by a hot IC medium.  A two-phase model, i.e.
cold clouds embedded in a hot IC medium, for the broad and narrow
line regions in quasars, BLR and NLR, respectively, has been
extensively studied (e.g. \shortciteNP{Krol81},
\shortciteNP{Mathews87}). The major objection to the model is the
cloud confinement problem (if the clouds are not continuously formed,
an external mechanism is necessary). Another problem is cloud break-up
by drag forces. The difficulties with the model, especially in
relation to the hot medium temperature, are discussed in
\shortciteN{Fabi86} and \shortciteN{Mathews87b}. The latter reference
discusses the dynamical implications of the two-phase model. Also in
\shortciteN{Elit86} the stability of these clouds is addressed.  Our
WA cloud model will be subject to all these caveats. The IC medium
could have a very significant effect on the dynamics of the clouds
through its drag and the resulting hydrodynamic instabilities.
Whether the WA clouds can move relative to a IC medium without
suffering gravitational, tidal, shear, or Rayleigh-Taylor
instabilities is beyond the scope of this work. We assume that
magnetic fields internal to the clouds must be important.

First we note that ignoring the drag force justifies hving chosen our
estimates as upper limits.  If the density of this external gas is
large enough then the WA would be subject to a drag force that would
cause us to overestimate the strength of gravity in the dynamics of
the WA. The upper-limit of the black hole mass is found by assuming
that the radiation force on the WA, $F_{\rm rad}$, is larger than the
gravitational restoring force, $F_{\rm grav}$. Therefore, the drag
force will be important if it is comparable to $F_{\rm rad}$:
\begin{equation}
\label{eq:equal}
{A L_{\rm abs} \over 4 \pi r^2 c} \approx \rho_{\rm 0} v^2 A.
\end{equation}
The density of the IC medium at which drag will be important can then
be related to the velocity of the WA through the medium (which we take
to be the observed radial velocity):
\begin{equation}
\label{eq:n_vs_v}
n_0 = {L_{abs} \over 4 \pi r^2 c \mu v^2},
\end{equation}
where $\mu$ is the mean mass of particles in the intercloud medium
(assumed to be a pure hydrogen and helium mixture). Making use of the
average $L_{\rm abs}$ for each object, we plot in Figure 6 this critical
density as a function of the outflowing velocity of the WA. A distance
of $r=10^{16}$~cm was assumed, and the range in $v$ was chosen to
cover the range that is typically observed in WAs.
\begin{figure}
\psfig{file=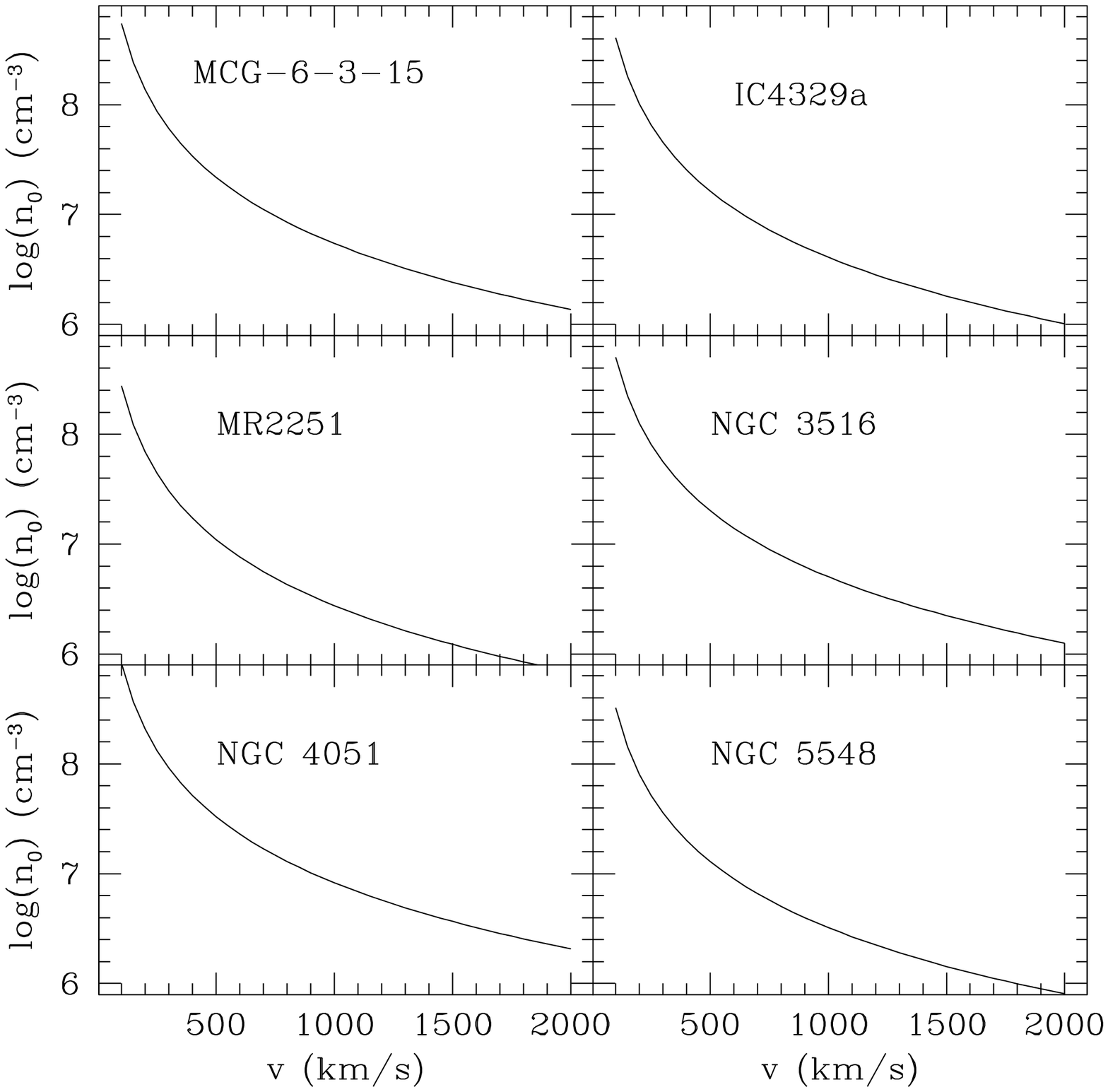,width=0.5\textwidth,silent=}
\caption{The density at which any external medium becomes dynamically
important is plotted as a function of the radial velocity of the
WA. For typical values of the velocity, a density $\approxgt
10^7$ \pcm is required for drag to have an effect. A distance of
$10^{16}$ \pcm for the WA has been assumed.}
\label{fig:nv}
\end{figure}
If the density of the IC medium is below the line then drag is not
important; alternatively, drag will be important if the IC density
lies above the line. For typical values of the observed WA blueshift,
an IC density $\approxgt 10^7$ \pcm seems to be required for drag to
be important in any of the objects. The most direct limits of the gas
come from opacity arguments (i.e. the absence of soft X-ray
absorption). We shall note at this stage that a medium with a density
of $10^7$ \pcm and a pathlength of $10^{16}$ \cm has a Thomson depth
of nearly 0.1.

Another constraint on the properties of the IC medium is that its
column density has to be smaller than that of the WA, $N_{\rm wa}$;
otherwise the IC medium could be identified with the WA itself.

\subsection{The acceleration term}
The acceleration term has to be positive for our estimates to
stand up as upper limits for the BH mass, and not lower limits as it
would be for the case of negative acceleration, or just measurements
of the BH mass for the case of zero acceleration. The dynamics of WA
clouds in pressure equilibrium with external magnetic pressure are
studied by \citeN{Chel01}. Their principal conclusion is that such
clouds can be accelerated to high velocities by means of radiation
pressure acceleration. The velocity profiles for two generic cases of
clouds (shell-like clouds with constant mass and constant density
clouds) are presented in their figure 10. For most external pressure
profiles, the acceleration results are positive for typical values of
WA velocities. Still for some cases a negative acceleration can be
possible.

Finally we would like to mention that for the WA temperature (around
around $10^5$\K), the plasma would be highly supersonic.  Therefore,
pressure forces would not be relevant for the velocities considered
(e.g. several hundred kilometers per second).

\section{DISCUSSION}
In this paper we present a new method for obtaining an upper limit on
an AGN BH mass by means of the comparison between the radiative and
gravitational forces acting on its WA. For that, we have adopted a
cloud model for the WA in which the radiation pressure dominates the
dynamics of such clouds. No further assumptions regarding the
geometry, location and nature of the WA are necessary. This method has
been applied to five Seyfert 1 galaxies (IC 4329a, MCG$-$6-30-15, NGC
3516, NGC 4051 and NGC 5548) and one radio-quiet quasar (MR 2251-178),
which cover a range of almost 4 orders of magnitude in X-ray
luminosity. The calculation of the upper limit of the BH mass required
the determination of only two quantities: the luminosity absorbed by
the WA cloud, $L_{\rm abs}$, and the column density of the WA, $N_{\rm
  wa}$. Both of them are obtained when the soft X-ray data is fitted
to our single zone photoionization models for the WA. Despite the
simplicity of our method, our estimates for the upper limit of the BH
mass fall within one order of magnitude of other independent methods
results (see Fig. 3).

The only requirement of our technique is the presence of outflowing
(i.e.  blueshifted) material which kinematics is dominated by
radiation pressure. UV absorbers could be used to determine the AGN BH
mass using the same method.  Narrow UV absorption features are shown
in many Type 1 AGN \shortcite{Cren99} and in most cases they appear
blueshifted with respect to the systemic velocity. Their typical
velocities are $\sim$ 1000 \kmps.

One uncertainty in our method is the presence of other non-radial
forces that affect the dynamic of the WA, and therefore change its
radial equation of motion (e.g. centrifugal or magnetic forces).
Further studies of the WA dynamics could possibly provide more
information on which forces are acting and therefore constrain more
tightly the determination of the BH mass. Also if the mass is
well-determined by some other means, application of our method can
lead to information on the drag and acceleration forces acting on the
WA.

Metallicities different than solar would have an effect on our
estimates. Given that $M_{\rm BH} \propto N_{\rm wa}^{-1} \propto
Z_{\rm wa}^{-1},$ where $Z_{\rm wa}$ is the metallicity of the warm
absorber, then solar and supersolar abundances maintain our estimates
as upper limits. It would be only in the case of subsolar abundances
when our upper limit could be lower than the real value of the black
hole mass. Note again that if the black hole mass can be accurately
determined, our method could be used to study the metallicity of the
warm absorber.

It is also worth noting that in the calculation of the absorbed
luminosity, $L_{\rm abs}$, we have used the fact that the WA is optically
thin material. This holds for the \asca data used in this paper (more
considerations on the determination of $L_{\rm abs}$ would be required if
we were to work with higher resolution data). Also better quality data
would lead to a more accurate determination on the errors of our upper
limits. 

One of the main goals of AGN studies is to measure the BH mass and a
myriad of techniques has been explored over the last two decades. The
use of WAs as vehicles to determine the AGN BH mass reinforces the
SMBH paradigm and opens up a powerful and simple technique for
measuring AGN BH masses.

\section{ACKNOWLEDGMENTS}
D.R. Ballantyne contributed with very useful discussions and careful
reading of the manuscript.  We want to thank Gary Ferland for
providing Cloudy and co-workers of P. Magdziarz for providing the
continuum in their paper \shortciteN{Magd98}. This work has been
supported by PPARC and Trinity College (R.M.) and by the Royal Society
(A.C.F.). This research has made use of the TARTARUS database, which
is supported by Jane Turner and Kirpal Nandra under NASA grants
NAG5-7385 and NAG5-7067.


\end{document}